# Direct observation of the spin texture in strongly correlated SmB$_6$ as evidence of the topological Kondo insulator

N. Xu,[1,]* P. K. Biswas,[2] J. H. Dil,[3,1] R. S. Dhaka,[1,3] G. Landolt,[4,1] S. Muff,[3,1] C. E. Matt,[1,5] X. Shi,[1,6] N. C. Plumb,[1] M. Radović,[1,7] E. Pomjakushina,[8] K. Conder,[8] A. Amato,[2] S.V. Borisenko[9], R. Yu[6], H.-M. Weng[6,10], Z. Fang[6,10], X. Dai[6,10], J. Mesot,[1,3,5] H. Ding,[6,10] and M. Shi[1],†

[1]*Swiss Light Source, Paul Scherrer Institut, CH-5232 Villigen PSI, Switzerland*

[2]*Laboratory for Muon Spin Spectroscopy, Paul Scherrer Institut, CH-5232 Villigen PSI, Switzerland*

[3]*Institute of Condensed Matter Physics, Ecole Polytechnique Fédérale de Lausanne, CH-1015 Lausanne, Switzerland*

[4]*Physik-Institut, Universität Zürich, Winterthurerstrauss 190, CH-8057 Zürich, Switzerland*

[5]*Laboratory for Solid State Physics, ETH Zürich, CH-8093 Zürich, Switzerland*

[6]*Beijing National Laboratory for Condensed Matter Physics and Institute of Physics, Chinese Academy of Sciences, Beijing 100190, China*

[7]*SwissFEL, Paul Scherrer Institut, CH-5232 Villigen PSI, Switzerland*

[8]*Laboratory for Developments and Methods, Paul Scherrer Institut, CH-5232 Villigen PSI, Switzerland*

[9]*Institute for Solid State Research, IFW Dresden, P. O. Box 270116, D-01171 Dresden, Germany*

[10]*Collaborative Innovation Center of Quantum Matter, Beijing, China*

**The concept of a topological Kondo insulator (TKI) has been brought forward as a new class of topological insulators in which non-trivial surface states reside in the bulk Kondo band gap at low temperature due to the strong spin-orbit coupling [1-3]. In contrast to other three-dimensional (3D) topological insulators (e.g. $Bi_2Se_3$), a TKI is truly insulating in the bulk [4]. Furthermore, strong electron correlations are present in the system, which may interact with the novel topological phase. Applying spin- and angle-resolved photoemission spectroscopy (SARPES) to the Kondo insulator $SmB_6$, a promising TKI candidate, we reveal that the surface states of $SmB_6$ are spin polarized, and the spin is locked to the crystal momentum. Counter-propagating states (i.e. at *k* and *-k*) have opposite spin polarizations protected by time-reversal symmetry. Together with the odd number of Fermi surfaces of surface states between the 4 time-reversal invariant momenta in the surface Brillouin zone [5], these findings prove, for the first time, that $SmB_6$ can host non-trivial topological surface states in a full insulating gap in the bulk stemming from the Kondo effect. Hence our experimental results establish that $SmB_6$ is the first realization of a 3D TKI. It can also serve as an ideal platform for the systematic study of the interplay between novel topological quantum states with emergent effects and competing order induced by strongly correlated electrons.**

Despite the great success in understanding the physical properties of materials proposed for topological insulators (TIs), a crucial experimental and technological problem remains unresolved – namely, most 3D topological insulator candidates are not bulk-insulating [4]. The bulk conducting band (BCB), as illustrated in Fig. 1b, makes it complicated to extract the topological surface properties from transport measurements, as well as for other bulk sensitive measurements. Furthermore, the bulk conductivity prohibits topological insulators from being applied as spintronic materials and from realizing exotic novel phenomena, such as Majorana fermions. Aside from this issue, at present topological insulators are essentially understood within non-interacting topological theory [6-7] (Fig. 1b), and thus the consequences of strong electron correlations interacting with topological phases are still unknown.

Recent theoretical investigations [1-3] have suggested that some Kondo insulators such as $SmB_6$, in which electrons are strongly correlated, can possibly host topological nontrivial surface states atop truly insulating bulk crystals, forming a special group of TIs known as topological Kondo insulators. Our previous high-resolution ARPES study [5] has revealed that the two dimensional surface states reside within the Kondo band gap ($\Delta \sim 20$ meV relative to chemical potential) and form three Fermi surfaces (FSs) in the first surface Brillouin zone (SBZ): one is centered at the SBZ ($\bar{\Gamma}$ point in Fig. 1a), and two (viewed as four half-Fermi surfaces) encircle the midpoints of the SBZ boundaries ($\bar{X}$ points) (Fig. 1c). The odd number of surface bands crossing the Fermi level ($E_F$) fulfills the necessary condition of topologically nontrivial surface states, and the observed surface state dispersions are in good agreement with theoretical calculations indicating that $SmB_6$ could be a topological Kondo insulator. Our observations are also consistent with transport results [8-15] and supported by other ARPES [16-18] studies, as well as STM measurements [19]. On the other hand, the in-gap states have also been discussed as topologically trivial polar surface states [20], and even as bulk states [21]. Until now, there has been no conclusive evidence indicating that $SmB_6$ is a TKI. Therefore, identifying the topologically non-trivial nature of the in-gap states of $SmB_6$ is crucial for distinguishing these different explanations and verifying whether SmB$_6$ is indeed the first realization of a TKI. Using spin-resolved ARPES, we investigated the spin texture of the surface states around the $\bar{X}$ points in the SBZ of SmB$_6$. In this letter we show that the surface states are spin polarized with strong $k_\parallel$ dependence. The spin-helical structure fulfills the requirement of time-reversal symmetry (TRS). Our results give direct evidence that SmB$_6$ is the first realization of a strongly correlated topological Kondo insulator.

In Figure 2b we plot the near-$E_F$ spin-integrated ARPES intensity and a momentum distribution curve (MDC) at $E_{SR}$ (10 meV above $E_F$) along the $\bar{X}$-$\bar{\Gamma}$-$\bar{X}$ direction. The momentum cut in the SBZ is indicated by the red line (C1) in the FS mapping (Fig. 2a). The spin-integrated ARPES data shown in Fig. 2a-2b are taken with $h\nu$ = 26 eV ($k_z$ = 0 for the bulk states) at the same experimental station as the spin-resolved ARPES data discussed in the following. The sample temperature for the spin-integrated and spin-resolved ARPES measurements is 20 K, at which the Kondo gap is fully opened and the surface states dominate near $E_F$ [5, 15-19]. Inside the Kondo band gap the α and β surface state bands are visible in the intensity plot, which are also clearly seen as peaks of the MDC at $E_{SR}$. Photon energy dependent measurements have shown that these bands do not disperse in $k_z$, the momentum along the surface normal, and thus are true surface states [5, 16-18]. To determine the spin polarization of the surface bands, we performed spin-resolved MDC measurements along the $\bar{X}$-$\bar{\Gamma}$-$\bar{X}$ cut (C1) at $E_{SR}$. The purpose of choosing the MDC to lie at $E_{SR}$ is to minimize the contribution of the bulk $f$ and $d$ states located at $E_B \geq$ 20 meV due to the limited energy resolution in spin-resolved mode (see Methods). At the same time, the count rate still allows us to acquire data with sufficient statistics in the SARPES measurements. Figures 2c, 2e and 2g show the spin resolved MDC intensity $I^{\uparrow\downarrow}_{x,y,z}$ in the $x$, $y$ and $z$ directions, respectively, measured with $h\nu$ = 26 eV and right-hand circularly polarized light (C+). The in-plane spin polarization $x$ and $y$ axes are defined along each $\bar{\Gamma}$-$\bar{X}$ direction ($x$ and $y$ as indicated in the coordinate system in Fig. 2a), and the out-of-plane spin polarization $z$ axis is along the sample normal $k_z$. As shown in Fig. 2c and 2d, there is a clear difference in $I^{\uparrow}_x$ and $I^{\downarrow}_x$ at each of the two MDC main peaks that correspond to two branches of the $\beta$ bands, indicating that the surface state bands are spin-polarized along the $x$ direction. We have also observed two more peaks between the main $\beta$ band peaks in the spin-polarization spectrum along the $x$ direction (Fig. 2d). These are ascribed to the $\beta'$ band (Fig. 1c), which results from the folding of $\beta$ due to a surface reconstruction assuming a single domain [5, 17]. We notice that the $\beta'$ band is hardly resolved from the $\beta$ band in the spin-integrated spectra (Fig. 2b) due to the limited resolution. However it can be observed clearly in the spin-resolved spectra because of the different spin polarization. Combining the spin-resolved ARPES data in Fig. 2, we can summarize that near $E_F$ the $\beta$ and $\beta'$ bands measured along C1 are both polarized in the $x$ direction, as indicated by the coordinate system in Fig. 2a. Moreover, each of the pairs of $\beta$ and $\beta'$ states located at ***k*** and -***k*** have opposite spin polarizations, consistent with the behavior of a spin-split Kramers pair, in accordance with time-reversal symmetry.

To determine the spin texture of the Fermi surface pockets, we have carried out spin-resolved ARPES measurements along the cuts C1-C4, as indicated in Fig. 3a, using C+ polarized light and photon energies of $hv$ = 26 eV and 30 eV. For both photon energies a consistent spin texture was obtained. As shown in Fig. 3c, the spin polarization (MDC at $E_{SR}$) measured with $hv$ = 30 eV along C1 is essentially the same as that in Fig. 2d taken with $hv$ = 26 eV. Figure 3d shows the spin polarization along C2 (perpendicular to C1). Similar to the observation along C1, the $\beta$ and $\beta'$ bands are spin polarized, but now along the $y$ direction. At the two $k_F$ points on cut C3 (C4), the spin polarizations of the $\beta$ band are opposite along the $x$ ($y$) direction (Figs. 3e and 3f). It should be mentioned that, internally consistent, the $\beta'$ peaks are not seen for contours C3 and C4 which do not pass through the $\beta'$ Fermi surface. The determined spin texture is summarized in Fig. 3a with spin polarizations marked by arrows. The measured spin texture, wherein the spin polarization is locked to the momentum, is fully consistent with topologically non-trivial surface states in the sense that it obeys both TRS and the crystal symmetry. This is further supported by the fact that the folded band *β'* has the same spin texture as the original band *β*, as expected from a simple Umklapp mechanism [22].

It has been shown that a spin polarization signal can also appear in the photoemission process from states that possess no net spin polarization [23]. This so-called photoemission effect is discussed, for instance, for the core-level photoelectrons from nonmagnetic solids [24], the bulk valence bands of topological insulators [25], as well as the bulk *f*-states of $SmB_6$ [26]. In contrast to the intrinsic spin signal from the spin polarized initial states, the non-intrinsic spin signal caused by the photoemission effect depends on the incident photon energy and polarization. For example, the non-intrinsic spin polarization of the *f*-states caused by the photoemission effect changes direction when the photon polarization changes from C+ to left-hand circularly polarized light (C-), and vanishes with linear polarization (*l-pol*) [26]. The consistent momentum-locked spin texture obtained with different photon energies (Figs. 2 and 3) in our experiments provides evidence that the observed spin polarizations are intrinsic to the spin polarized initial states. To rule out the possibility that the detected spin texture results from the polarization of the incident light, we conducted spin-resolved measurements using incident light with all available polarizations. Figures 3f, 3g and 3i show the results along C4 detected by using C+, C-, and linear polarizations of the incident light. The same spin polarizations from differently polarized light give further confidence that the observed spin polarizations reflect the intrinsic spin structure of the initial states.

The results obtained with the linearly polarized light (Figs. 3h and 3i) also indicate that the spin polarization originates from the surface the *β* band, and not from contamination of the bulk *f* states due to the limited energy resolution used in SAPRES measurements because spin polarization of the bulk *f* states caused by the photoemission effect should vanish with linearly polarized light [26]. To further exclude contamination from the bulk *d*-states as an origin of the observed spin polarizations of the MDCs studied at $E_{SR}$, we performed measurements at a higher binding energy ($E_{HB}$ as illustrated in Fig. 3b), where the bulk *d*-states dominate the photoemission intensity. Figures 3j and 3k show the spin resolved MDC intensity $I^{\uparrow\downarrow}$ and the spin polarization spectra along the *y* direction, taken with linearly polarized light (*l-pol*). In contrast to the spectra at $E_{SR}$ (Figs. 3h and 3i), the spin-resolved MDCs at $E_{HB}$ show negligible difference, indicating that the photoelectrons from the spin-degenerate bulk *d*-states are not spin polarized. Therefore we conclude that the spin signal at $E_{SR}$ is dominated by the surface states, which provides compelling evidence that $SmB_6$ is the first realization of a TKI.

Figure 4 schematically summarizes our main experimental finding of the spin structure of the surface bands of $SmB_6$. For simplicity, we consider the situation without folding bands. The spin orientations of the electrons in the in-gap surface *β* band, which is located between the strongly localized bulk *f*-states and the chemical potential, are locked to their momenta; namely, at opposite momenta (*k* and -*k*), the surface states have opposite spins. This anti-clockwise spin texture for the surface band in $SmB_6$ shows a great similarity to other three-dimensional TIs ($Bi_2Se_3$, $Bi_2Te_3$ etc.) which each has a FS from a single pocket centered at the $\bar{\Gamma}$ point in the SBZ [27-28]. Extensive ARPES studies have shown that $SmB_6$ has a surface FS formed by 3 electron-pockets with Kramers points located around the SBZ center and boundary [5, 16-17]. The revealed spin texture, together with the odd number of pockets, indicates that the metallic states at the surface of $SmB_6$ are non-trivial topological surface states and confirm recent theoretical predications [1-3]. A crucial and so far unresolved problem for the real-world implementation of TIs is that most 3D topological insulator candidates are not bulk-insulating [4, 29]. On the other hand, $SmB_6$ is a mixed valence Kondo insulator. Due to the hybridization of the nearly localized *4f* bands with the dispersive conduction band, a band gap opens near the chemical potential at low temperature, leading the system in to a good bulk insulator [4, 30-32]. Thus the direct identification of $SmB_6$ as a topological Kondo insulator is a significant advance in realizing a topological quantum state of matter in which the metallic edge states, protected by TRS, are located on a true bulk insulator. It is also worthwhile to mention that the metallic non-trivial surface states formed by the electron-pockets at low temperatures can provide a natural explanation for the longstanding puzzle that the resistivity in SmB$_6$ saturates to a finite value instead of being divergent [30-32]. On the other hand, the coexistence of the topological surface states and the strongly correlated bulk states offers new opportunities for understanding topological insulators beyond the non-interacting topological theory.

## Methods:

**(S)ARPES measurements.**

Spin- and angle-resolved photoemission spectroscopy measurements were performed at the Surface/Interface Spectroscopy beamline at the Swiss Light Source with the COPHEE station. The two orthogonally mounted 40 kV classical Mott detectors allow for the determination of all three components of the spin of the electron ($P_x$, $P_y$ and $P_z$) as a function of its energy and momentum [33-34]. To achieve this the spin polarization curves obtained for the three spatial directions are fitted simultaneously with the sum of all the Mott channels; i.e. the spin integrated signal. Further technical details of the data acquisition and analysis can be found in [35]. The SARPES measurements were repeated for several samples with photon energies ranging from 26 to 30 eV and with circular and linear light polarizations. Results with different photon energies and light polarizations are consistent. All the spin-integrated and spin-resolved data shown in Figs. 2 and 3 are taken at $T = 20$ K, where the Kondo gap is fully open [7,15-18]. The energy and angle resolutions were 60 meV (FWHM) and 3% of the SBZ, respectively, to allow for enough statistics. Because of the low photoemission signal, the typical duration of a spin-resolved MDC was about 6 hours. After this period the sample quality was confirmed again by spin-integrated measurements. Clean surfaces for the (S)ARPES measurements were obtained by cleaving the crystals *in situ* at low temperature (20 K) in a working vacuum better than $2 \times 10^{-10}$ mbar. Shiny mirror-like surfaces were obtained after cleaving the samples, confirming their high quality.

**Sample fabrication.**

High-quality single crystals of $SmB_6$ were obtained by growth from Al-flux using samarium pieces (99.9%), boron powder (99.99%) and aluminum pieces in a ratio of 0.5 g of Sm:B (1:6) and 50g of Al (99.99%). The growth was done in a vertical gradient furnace under continues argon flow at a temperature up to 1500 °C with a cooling rate 5 K/hour. The aluminum flux was removed by potassium hydroxide solution. The crystals, which are millimeter-size, have a bar shapes, black color, and flat, mirror-like surfaces.

## Acknowledgement:

This work was supported by the Sino-Swiss Science and Technology Cooperation (Project No. IZLCZ2138954), the Swiss National Science Foundation (Grant No. 200021-137783 and PP00P2_144742), and MOST (Grant No. 2010CB923000) and NSFC.

## Author contributions:

N.X., J.H.D. and M.S. conceived the experiments. N.X., J.H.D., R.S.D., G.L., S.M., X.S. carried out the experiments with assistance from C.E.M., N.C.P., M.R., S.V.B.; N.X., J.H.D. and M.S. analyzed the data with valuable feedback from R.Y., H.M.W., Z.F., X.D., J.M. and H.D.; P.K.B., E.P., K.C., and A.A. synthesized the samples; N.X. and M.S. wrote the manuscript. All authors discussed the results and commented on the manuscript.

## Additional information:

The authors declare no competing financial interests.

**Correspondence** and requests for materials should be addressed to N.X. (e-mail: nan.xu@psi.ch) or M.S. (e-mail: ming.shi@psi.ch).

**Figure Legends:**

**Figure 1:** (a) The first Brillouin zone of $SmB_6$, and the projection on the cleaved surface. High-symmetry points are also indicated. (b) Illustration showing the electronic correlation and bulk insulating nature of $Bi_2Se_3$ and the topological Kondo insulator $SmB_6$. The inset shows the electronic structure and spin texture of $Bi_2Se_3$. (c) 3D intensity plot of the photoemission data, showing the Fermi surface and electronic structure of $SmB_6$ measured with a photon energy $h\nu = 26$ eV and energy resolution < 5 meV at $T = 1$K.

**Figure 2: Spin polarization of topological Dirac surface states along the $\bar{X}$-$\bar{\Gamma}$-$\bar{X}$ direction.** (a) Fermi surface map measured with $h\nu = 26$ eV. (b) Low energy excitations near $E_F$ along the high symmetry line $\bar{X}$-$\bar{\Gamma}$-$\bar{X}$. The location in the $k_x$-$k_y$ plane is indicated by C1 in (a). The red curve is the momentum distribution curve at $E_{SR}$ (10 meV above $E_F$). (c) Measured spin-resolved intensity projected on the $x$ direction for the surface states at $E_{SR}$. The red and blue symbols are the intensity of spin up and spin down states, respectively. (d) Spin polarization along the $x$ direction for the surface states measured at the $E_{SR}$. (e)-(f) Same as (c)-(d) but along the $y$ direction. (g)-(h) Same as (c)-(d) but along the $z$ (out-of-plane) direction.

**Figure 3: Spin polarization of topological Dirac surface states along high symmetry lines.** (a) Schematic showing the Fermi surface of the topological surface states in $SmB_6$, with red lines indicating the locations of spin measurements in the $k_x$-$k_y$ plane, labeled as C1, C2, C3 and C4. The circle at the BZ center $\bar{\Gamma}$ point and ellipses at the middle points of BZ boundaries ($\bar{X}$ points) indicate the Fermi surfaces of the α and β bands, respectively [5, 17]. The dashed line ellipse at the $\bar{\Gamma}$ point indicates the folded band β' [5, 17]. The dots are the $k_F$ positions of the bands, and the arrows are the measured spin polarizations at the positions of the dots. The black dashed lines indicate the first surface Brillouin zone. (b) Low-energy excitations near

$E_F$ along the high symmetry line $\bar{M}$-$\bar{X}$-$\bar{M}$, measured with photon energy $h\nu$ = 30 eV. The location in the $k_x$-$k_y$ plane is indicated by C4 in (a). (c) Spin polarization measured at $h\nu$ = 30 eV along the *x* direction for the surface states at $E_{SR}$ (10 meV above $E_F$.) The location in the $k_x$-$k_y$ plane is indicated by C1 in (a). (d) Analogous to (c), but for C2, where the spin polarization direction is along *y*. (e) Analogous to (c), but for C3, where the spin polarization direction is along *x*. (f) Analogous to (c), but for C4, where the spin polarization direction is along *y*. For (c)-(f) the photon polarization is right hand circular (C+). (g) Same as (f), but with C- photon polarization. (h) Measured spin-resolved intensity along the *y* direction for the surface states at $E_{SR}$, measured at $h\nu$ = 30 eV with linear light polarization. The red and blue symbols are the intensity of spin up and spin down states, respectively. The location in the $k_x$-$k_y$ plane is indicated by C4 in (a). (i) Spin polarization along the *y* direction for the surface states at the $E_{SP}$. (j)-(k) Same as (h)-(i) but at higher binding energy illustrated by $E_{HB}$ in (b).

**Figure 4: Schematic of the spin-polarized surface state dispersion in the topological Kondo insulator SmB$_6$.** The spin polarizations are indicated with red arrows. The shape and size of the α band are from references [5, 17].

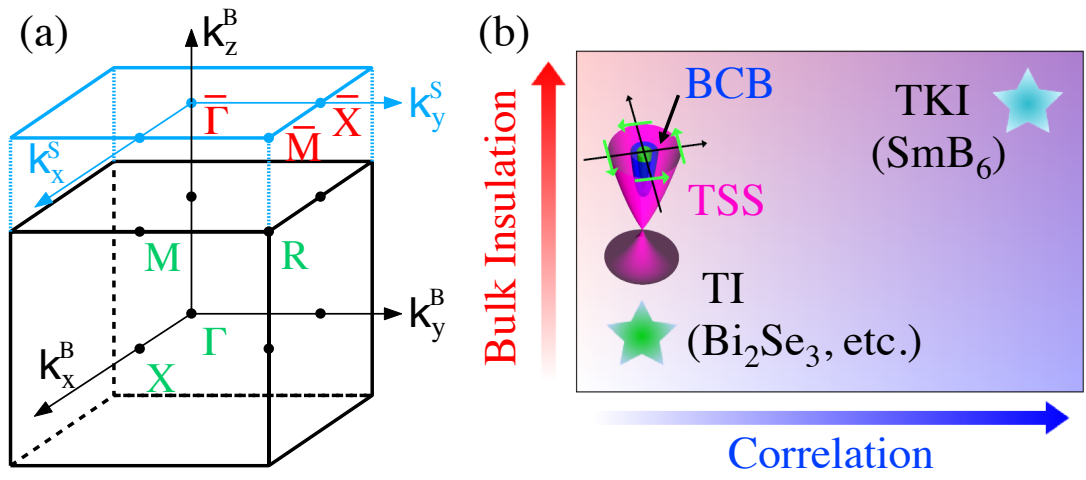
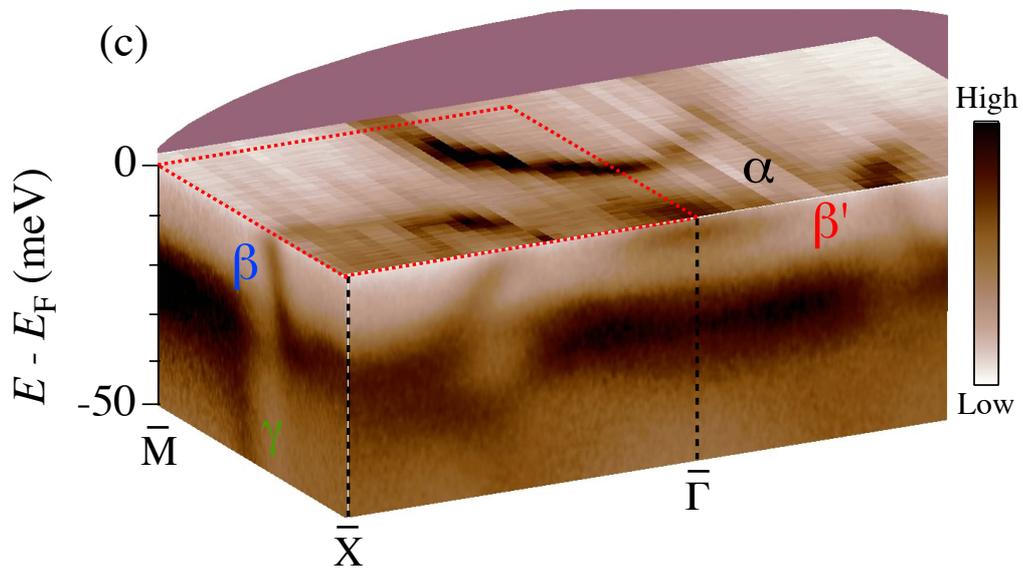

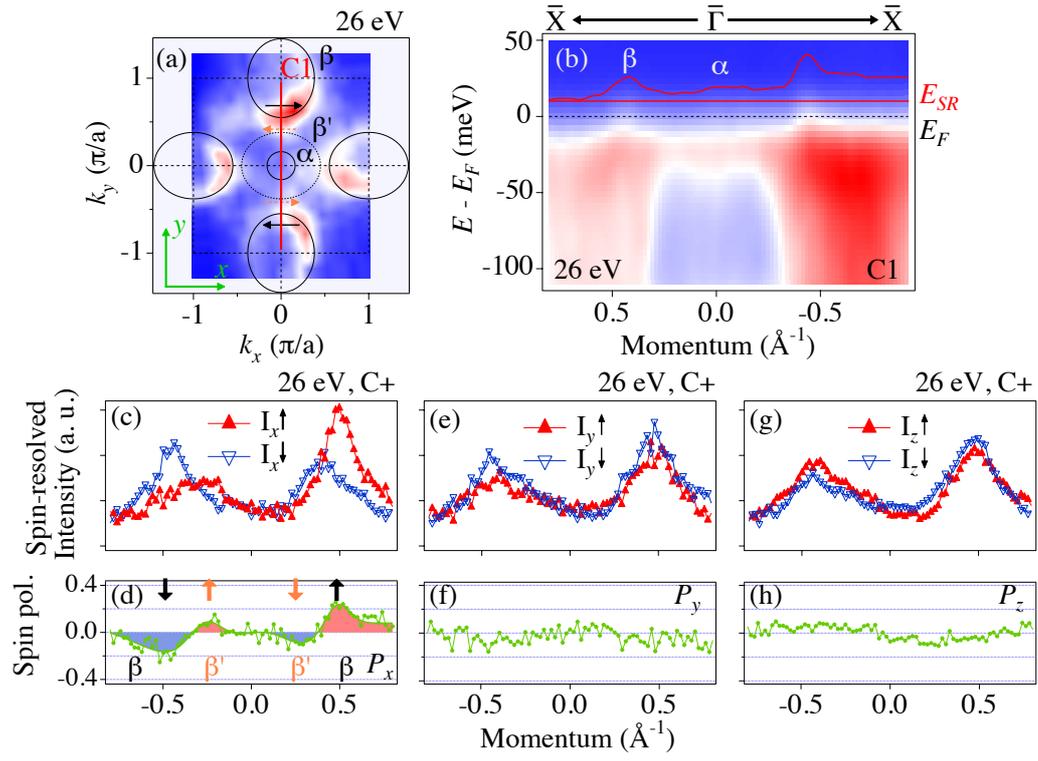

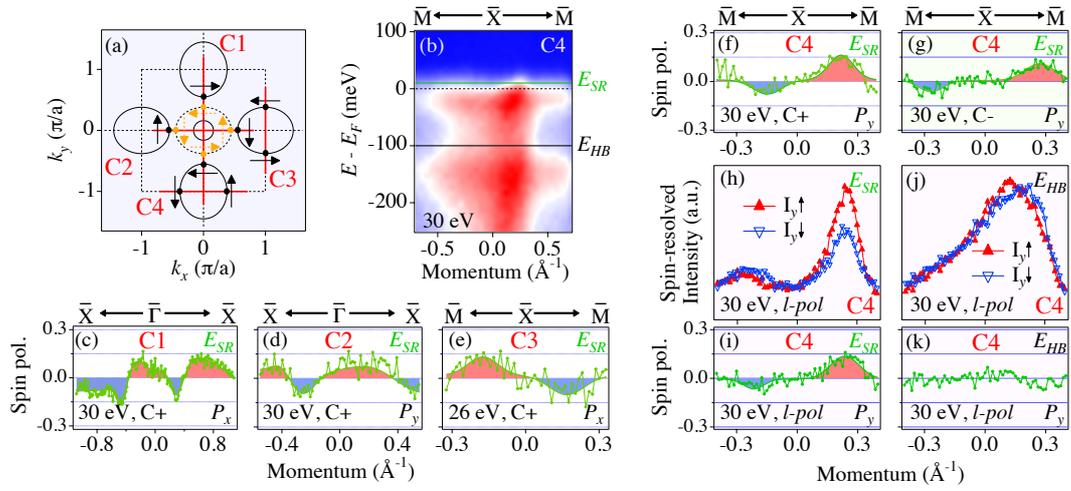

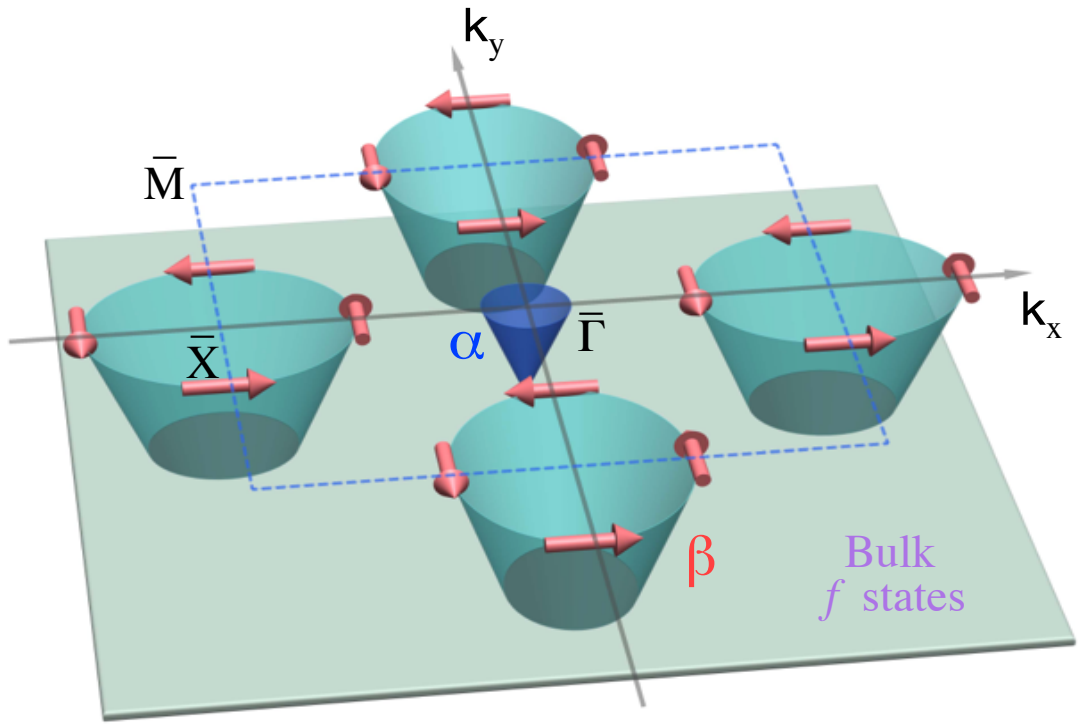